# Generic visuality of war? How image-generative AI models (mis)represent Russia's war against Ukraine

Mykola Makhortykh and Miglė Bareikytė

**Abstract:** The rise of generative AI (genAI) can transform the representation of different aspects of social reality, including modern wars. While scholarship has largely focused on the military applications of AI, the growing adoption of genAI technologies may have major implications for how wars are portrayed, remembered, and interpreted. A few initial scholarly inquiries highlight the risks of genAI in this context, specifically regarding its potential to distort the representation of mass violence, particularly by sanitising and homogenising it. However, little is known about how genAI representation practices vary between different episodes of violence portrayed by Western and non-Western genAI models. Using the Russian aggression against Ukraine as a case study, we audit how two image-generative models, the US-based Midjourney and the Russia-based Kandinsky, represent both fictional and factual episodes of the war. We then analyse the models' responsiveness to the war-related prompts, together with the aesthetic and content-based aspects of the resulting images. Our findings highlight that contextual factors lead to variation in the representation of war, both between models and within the outputs of the same model. However, there are some consistent patterns of representation that may contribute to the homogenization of war aesthetics.

**Funding statement**: Gefördert durch die Deutsche Forschungsgemeinschaft (DFG) – Projektnummer 262513311 – SFB 1187 (Funded by the Deutsche Forschungsgemeinschaft (DFG, German Research Foundation) – Project-ID 262513311 – SFB 1187). The Alfred Landecker Foundation provided financial support for the research time of Dr. Makhortykh.

# Generic visuality of war? How image-generative AI models (mis)represent Russia's war against Ukraine

## Introduction

The rise of generative artificial intelligence (genAI), a technology that allows "generating seemingly new, meaningful content such as text, images, or audio from training data" (Feuerriegel et al., 2024, p. 111), may transform how social reality is (mis)represented. GenAI facilitates and accelerates the production of multimodal artificial content that imitates the outcomes of (often time- and resource-demanding) human labour and representation practices, from artwork to journalistic reports. The emerging genAI industries may affect both the representation of specific issues - e.g. by decreasing the diversity due to homogenising tendencies of genAI (Karell et al., 2025) or propagating false or distorted interpretations of reality (Makhortykh et al., 2023) - and the potential effects of such representation - e.g. by decreasing epistemic trust among AI users (Sahebi & Formosa, 2025) or shifting their opinion on a specific issue that is represented (Costello et al., 2024).

One area in which such genAI-driven developments raise particularly pronounced concerns is the representation of modern wars. While there is a rapidly growing interest in the relationship between AI technologies and modern warfare (Jensen et al., 2020), most of existing research focuses on the military applications of AI, ranging from the enhancement of autonomous drone performance (Kunertova, 2023; Makhortykh & Bareikytė, 2025) to the ethical mechanisms of ensuring human control over the autonomous weapons (Amoroso & Tamburrini, 2020; Umbrello, 2021). However, genAI also shapes how wars are portrayed and communicated about, both by human actors who have traditionally been involved in representing mass violence (e.g. journalists or propagandists) and non-human actors, such as AI chatbots.

Image-generative AI models are particularly relevant in this context due to the power of images, especially photographs, to communicate meanings, transmit affect, and form memories (Sontag, 2003). Concomitantly, visual representations of war by genAI are shaped by probabilistic mechanisms based on existing representations, which serve as training data for AI models. It results in the risk of genAI industries producing "highly homogenized, regimented, and proceduralized" (Laba, 2024, p. 1600) visions of mass violence, raising the question of whether genAI can enable meaningful representation of mass violence (Gillespie, 2024). A risk of homogenization is amplified by the inability of genAI models to differentiate between ethically appropriate and inappropriate outcomes (unless a thorough system of safeguards is implemented), resulting in the possibility of them facilitating the creation of distorted or fictional, and factually incorrect, representations of mass violence.

The importance of understanding how genAI can influence the visual representation of modern wars is particularly pronounced in the context of the Russian aggression against Ukraine. Not only is it a long-lasting and intense war, but it also coincides with a series of radical changes in media technologies relevant for war representation (Ford & Hoskins, 2022; Ford, 2025). The immediate consequence of such a change is the increased visibility of the war in Ukraine outside the country, due to more possibilities for its digital coverage and witnessing, as well as intensified contestation of its representation. Since the beginning of the war in 2014, its mediatization has been characterised by the extensive amount of disinformation and propaganda (Stratcom, 2015), which has further increased following the large-scale invasion in 2022 (OECD, 2022). Against this backdrop, the visuality of the war produced by genAI may play an important role in shaping the already contested representation of Russia's consequences, for instance, in war remembrance (Laba et al., 2025).

Until now, most research on the visual representation of Russia's war in Ukraine by genAI has focused on the amplification of disinformation and propaganda (Tolmach et al., 2024; Twomey et al., 2023). Only a few studies (Laba, 2024; Laba et al., 2025) have considered the broader implications of genAI for war representation, noting its tendency to stereotypise the war portrayal, resulting in "generic, decontextualised, and often dehumanised depictions of conflict" (Laba et al., 2025, p. 17). As argued by Aris et al. (2023), human-machine cooperation in visual representations and digital art has a long history, involving the development of digital sketching devices, computer interfaces, and software. In this context, GenAI can potentially enhance human creativity; however, it also lacks the emotional, experiential, and socio-cultural depth that is crucial for human-made representations (Aris et al., 2023). Instead, GenAI models and predicts societal aesthetic preferences, including the automatised practices of beautification or blurring, raising the question of aesthetic diversity (Manovich, 2018). An immediate consequence of it for war representation can be the loss of contextual specificity and reiteration of a few representation tropes; however, it remains unclear to what degree such processes may be affected by the data and fine-tuning practices of genAI models and different user prompts (e.g. regarding actual and fictional episodes of the war).

To address this uncertainty, we conduct a comparative audit of the Western (Midjourney) and Russian (Kandinsky) genAI models to examine how homogeneous their representation of both factual and fictional aspects of Russia's war in Ukraine is. To do so, we first discuss how different forms of media technology influence war representation, followed by a brief review of research on the mediatization of Russia's war in Ukraine. We then introduce the methodology we used to conduct the audits of Midjourney and Kandinsky, along with our data analysis strategy. Following this, we present our findings regarding the responsiveness of the models to war-related prompts, the presence of specific content elements related to representation, and its aesthetic aspects concerning the image perspective and the distribution of light. We conclude with a discussion of the implications of our findings for research on war representation, together with the limitations of the current study and directions for future research.

**War representation in the age of genAI**

Different forms of media have long been used to represent wars. The origins of mediated war representation are often traced back to the use of photography to represent armed conflicts in the 19th century (e.g. the Crimean War; Hüppauf, 1993). The 20th century accelerated this process with new forms of media, notably radio broadcasting, film, and television, and the intensification of mass violence, including two world wars (Sontag, 2003). The unprecedented amount of representations of war and associated violence, along with their media-enabled visual and auditory realism and extensive reach, turned violence into a form of "mass entertainment" (Boichak & Hoskins, 2022, p. 4) and increased individual engagement with it. The increased exposure to information about the wars had multiple consequences, ranging from public mobilisation and rallying around the flag (Kizilova & Norri, 2024) to the erosion of trust in governments involved in the conflicts and the rise of anti-war sentiments (Pickerill & Webster, 2006).

The growing adoption of digital media signified an important shift in how wars are represented. The affordances of digital platforms like TikTok, Telegram or YouTube enable new multimodal forms of representation, ranging from amateur video recordings to news reports to the use of digital media representations for the military practice of targeting and killing, also known as the kill chain (Smith & McDonald, 2011; Ruchel-Stockmanns, 2018; Divon & Eriksson Krutrök, 2024; Ford, 2025). The accessibility of these affordances expands the range of individuals involved in war representation as witnesses, reporters, commentators, or propagandists. It makes the experiences of war feel even more immediate than those of analogue media, increasing global awareness about the humanitarian costs of violence and facilitating public mobilisation to resist violence (Boichak & Hoskins, 2022). However, this change also amplifies the spread of misleading and propagandist content that can contribute to violence or foster cynical attitudes towards suffering (Shields, 2021; Boichak & Hoskins, 2022).

The increasing accessibility of genAI brings further changes in how modern wars are represented and engaged with. The artificial representations of mass violence, varying from genAI-made texts to videos, are embedded in today's datafied (visual) cultures (Bareikytė & Skop, 2022), where human-generated data (including representations of violence and suffering) is used to train AI. GenAI does not provide yet another tool to represent wars in a more realistic manner, akin to a sharper camera lens, but rather destabilises the human-centred act of war representation. Such destabilisation is due to genAI models relying on the probabilistic appropriation of datafied human labour, both regarding training data acquisition and its processing (Hao, 2025), while failing to capture meaning in a human sense (Wasielewski, 2023). However, while genAI models do not “understand” or “see” war in the same way humans do, their outputs can still affect people engaging with them. While not inherently harmful, such engagement raises concerns about the risk of AI reiterating societal biases (Benjamin, 2019) or propagating disinformation (Wack et al., 2025).

To date, much research on AI and representations has focused on how humans imagine AI (Paltieli, 2022; Hansen, 2022; Linderoth et al., 2024). In contrast, we examine what representations are created by AI models and the potential homogenising tendencies in their visual output. This is urgent in the context of war representations, because digital (and increasingly AI-driven) media shape public perception of violence through visual information and disinformation (Bareikyte & Skop, 2022; Ford, 2025). Wars historically have been shaped not only by the situation at the frontline but also by how violence is imagined, making representation another “medium of combat” (Mieszkowski, 2012). With human understanding of specific wars being affected by genAI, artificial representations of violence become both epistemic tools and hybrid weapons. To better understand this ambiguity, we examine how genAI models represent Russia’s war in Ukraine and how their war gaze differs from other digitally mediated forms of representation.

**Digital media representations of Russia’s war in Ukraine**

Since its beginning in 2014, Russia’s war against Ukraine has been actively represented in (digital) media and has become the subject of multiple studies looking at the relationship between digital technologies, associated user practices, and war representation. Until now, most studies have focused on non-AI forms of representation, such as journalistic reporting (Nygren et al., 2018; Midões & Martins, 2023; Schumacher et al., 2024) or social media practices (Makhortykh & Lyebyedyev, 2015; Bösch & Divon, 2024; Bareikytė & Makhortykh, 2025). A separate strand of research examines the disinformation and propaganda used to contest and distort the representation of the war as part of the Kremlin’s information warfare, which spans multiple digital platforms (Mejias & Vokuev, 2017; Tolz & Hutchings, 2023; Gaufman, 2023).

The existing research highlights the diversity of forms and purposes that these cross-platform representation practices serve. Some of them are used to mobilise support for the war and incite violence and hate (Gaufman, 2015; Makhortykh, 2018), whereas others serve the purpose of stimulating the discussion about violence (Wiggins, 2016), mourning the casualties (Urman & Makhortykh, 2025a) or documenting the events of the war (Bareikytė et al., 2024). Several of these studies (e.g. Makhortykh & Sydorova, 2017; Tao & Peng, 2023; Oleinik, 2025) adopt a comparative angle, noting pronounced differences in cross-country war representation, which prompts our interest in comparing Russian and Western AI models.

With the development of AI, digital media affordances become increasingly shaped by the automated systems for organising, prioritising, and, in some cases, personalising content. The adoption of these automated content curation systems has implications for war representation, for instance, by promoting or undermining journalistic practices in war reporting (Makhortykh & Bastian, 2022) or by placing individuals in information bubbles associated with specific perspectives on Russia's war (Hamidy, 2024). Simultaneously, the increased capacities for content production and distribution, specifically (ro)bots, raise concerns about new possibilities for manipulating war-related representations (Smart et al., 2022; Zhao et al., 2024).

The rise of genAI prompted a new wave of scholarly interest in its possible effects on the mediatization of Russia's war in Ukraine. Several studies (Makhortykh et al., 2024; 2025; Alyukov et al., 2025) examined how AI applications integrated with search engines produce textual content about the war and found a troubling tendency for some applications to reiterate claims associated with the Kremlin propaganda, together with the variation in war representation due to the stochasticity and localisation. A few studies (Laba, 2024; Laba et al., 2025) looked specifically at how image-generative AI represents the war, finding the tendency for some models, particularly Midjourney, to homogenise the war representation, resulting in the prevalence of a few visual tropes (e.g. fighters and destruction; Laba, 2024).

## Methodology

### *Data collection*

To conduct the study, we employed AI auditing, a research technique that systematically examines the performance of AI models and applications to understand their functionality and impact (Kuznetsova et al., 2025). We manually audited two popular genAI models: Midjourney and Kandinsky. Midjourney was developed by a US-based company, Midjourney, Inc., using the public LAION-5b dataset (Batt, 2025). The dataset relies on a snapshot of the Common Crawl repository and is more likely to include offensive and illegal materials (e.g. related to child sexual abuse; Thiel, 2023); however, it is also more transparent than proprietary datasets used by specific AI companies (Schuhmann et al., 2022). By contrast, Kandinsky was developed by the information technology team at Sber, a majority Russian state-owned bank, using the CLIP dataset (Razzhigaev et al., 2023), which is proprietary and originates from OpenAI. At the time of our study, to access Midjourney, users required paid access via a Discord bot, whereas Kandinsky was available for free via a Telegram bot.

For the audit, we focused on the representation of Russia's war in urban environments in Ukraine. This focus is attributed to the predominantly urbanised nature of warfare since the full-scale Russian invasion, which is a distinct feature of the war (King, 2024). We used the following prompt structure to generate data: “Generate a professional photo of the city [A] [under condition B] [inflicted by party C]". The first part of the prompt remained stable for consistency and included magic words - "a professional photo" - which are the terms expected to improve the performance of genAI models by triggering “patterns that have been encoded in the model” (O’Reilly, 2012). It also included the expected use case, i.e. urban environments affected by the war. Here is an example of how the prompt eventually appeared: “Generate a professional photo of the city of Avdiivka under attack by Russia.”

For A, we used the list of 41 cities/towns in Ukraine which were either a) destroyed and occupied by the Russian army; b) occupied but mostly undamaged; c) damaged but not occupied; d) relatively little affected by the war in 2024. The list was compiled by the authors based on Bellingcat data on the urban destruction in Ukraine (https://ukraine.bellingcat.com). The complete list of cities, along with their corresponding categories, is provided in Appendix 1. For B, we used three conditions: a) under attack; b) occupied; and c) liberated; the selection was due to our interest in whether AI representations will vary across the prompt-specific conditions. For C, we initially planned to use three parties - Russia, Ukraine, and NATO - which are in one way or another involved in the war. The NATO option was included to examine how the models would perform for prompts inquiring about fictional representations: while NATO is not attacking, occupying, or liberating any Ukrainian cities, it is part of the Kremlin's disinformation, which could poison AI training data. However, because Kandinsky declined to generate outputs for prompts that included the word “Ukraine”, we included only Russia- and NATO-focused prompts.

We generated data in the first two weeks of March 2024. It is important to note that, as genAI models are constantly being developed, our results may not be representative of the current state of the models. Instead, we captured the historical perspective on how models represent war, which evolves due to the changes in the models’ architecture, safeguards, or the users’ access to the models. These changes can also be influenced by the developments at the frontline, as the Russian aggression against Ukraine continues, and specific cities can still be occupied or freed, which, in turn, can impact their AI representations. While this historical perspective is a limitation of the study, it also provides an important benchmark for understanding how AI representations of the war change over time.

*Data analysis*

To analyse the outputs of genAI models, we employed two approaches. First, we applied descriptive statistics to examine the models' responsiveness to war-related prompts. Specifically, we were interested in whether models generate the content requested by the user (including images in response to requests for fictional or ideologically skewed content, such as the occupation of Ukrainian cities by NATO or the so-called “liberation” by Russian troops) or if prompts are consistently blocked by the safeguards (for instance, due to pro-Kremlin censorship; Urman & Makhortykh, 2025b).

Second, we used qualitative content analysis to examine certain aesthetic aspects of war represented by genAI models. We specifically focused on three aspects of aesthetic war representation, which we found particularly relevant in terms of how mass violence can be perceived. First, we examined the visual content elements included in the images (e.g. humans, ruination, or religious symbols), using as a starting point the list of visual elements commonly found in representations of Russia's war in Ukraine on social media (Makhortykh & Sydorova, 2017). To label this aspect of representation, we manually coded content elements present on the images (for the complete list, see Figure 3 in the Findings). Further, we examined the aesthetic aspects by focusing on a spatial perspective of the image and the distribution of light. Based on the inductive close reading, we differentiated between the two most common spatial perspectives: the top-down (or the drone) and the human perspective. During the coding process, a third—i.e. in-between—category was added for images where it was difficult to discern whether the perspective was that of a drone or a person standing in a high-rise building. For the distribution of light, we differentiated between artificial (e.g. from lamps) and natural lighting (e.g. daylight or moonlight). After the initial coding, these two variables were expanded, resulting in the final set of lighting-related categories: 1) natural-gloomy: daylight images with clouds, but no artificial lighting or direct sun rays; 2) natural-sunny-cloudy: daylight images with direct sun rays and clouds, but no artificial lighting; 3) natural-sunny: daylight images without clouds or artificial light; 4) artificial-night: nighttime images with artificial lighting; 5) artificial-night-sunny: nightlight images with sunlight and artificial lighting at the same time (due to the image being produced by a potentially hallucinating genAI); 6) natural-gloomy-sunny: daylight images with clouds and very few sunlight or sun rays.

## Findings

### *Responsiveness of models to the prompts concerning Russia's war in Ukraine*

We started our analysis by examining how responsive Midjourney and Kandinsky were to the war-related prompts. Figure 1 shows that Midjourney generated outputs for all the prompts, including those asking for fictional episodes of the war, such as the attacks on Ukrainian cities by NATO or the NATO occupation of Ukraine. These fictional cases also included the occupation of Ukrainian cities by Russian forces, which never happened in reality, for instance, of Kyiv, Lviv, or Kharkiv. Similarly, Midjourney generated content for prompts that opposed Russian official discourse (e.g. by referring to the capture of Ukrainian cities as an occupation) and aligned with it (e.g. for prompts referring to the "liberation" by Russia).

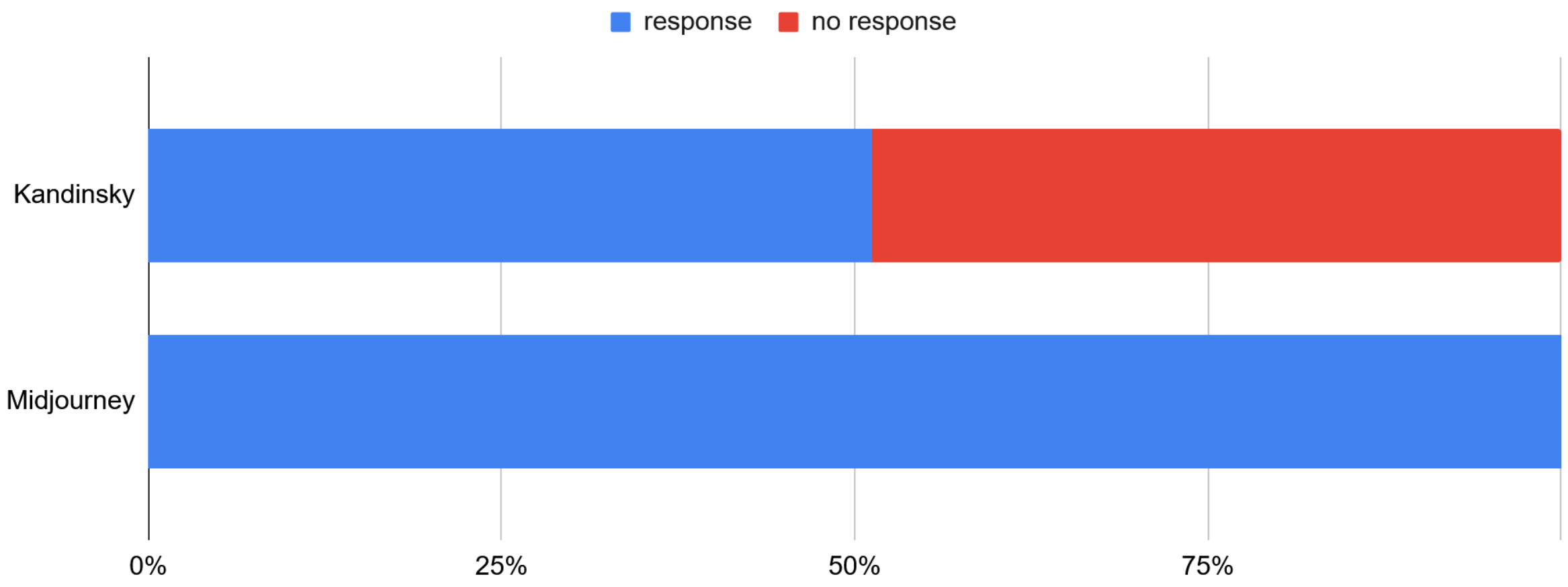

**Figure 1**. Responsiveness of Kandinsky and Midjourney to war-related prompts

By contrast, Kandinsky provided outputs only to 51% of prompts, whereas 49% of prompts triggered the safeguards, which prohibited image generation. Together with Kandinsky's refusal to generate content for prompts with the word “Ukraine”, which we noted earlier, it suggests that the Russia-made model is substantially more prone to censorship regarding war representation. To explore in more detail which prompts are particularly likely to trigger safeguards, we mapped the distribution of no responses across specific prompts grouped by the conditions, actors involved in inflicting these conditions, and groups of cities (Figure 2).

This additional analysis reveals that safeguards were triggered particularly frequently for prompts related to Russia's attacks on Ukrainian urban spaces: 100% of all prompts resulted in no responses. Surprisingly, there was also censorship of images showing the so-called "liberation" by Russia, despite such prompts aligning with the Kremlin’s propaganda. Kandinsky was particularly hesitant to provide responses to prompts mentioning the attack; however, while it could, to a certain degree, be attributed to the model trying to prevent the generation of violent content, it did generate some images when the actor was NATO.

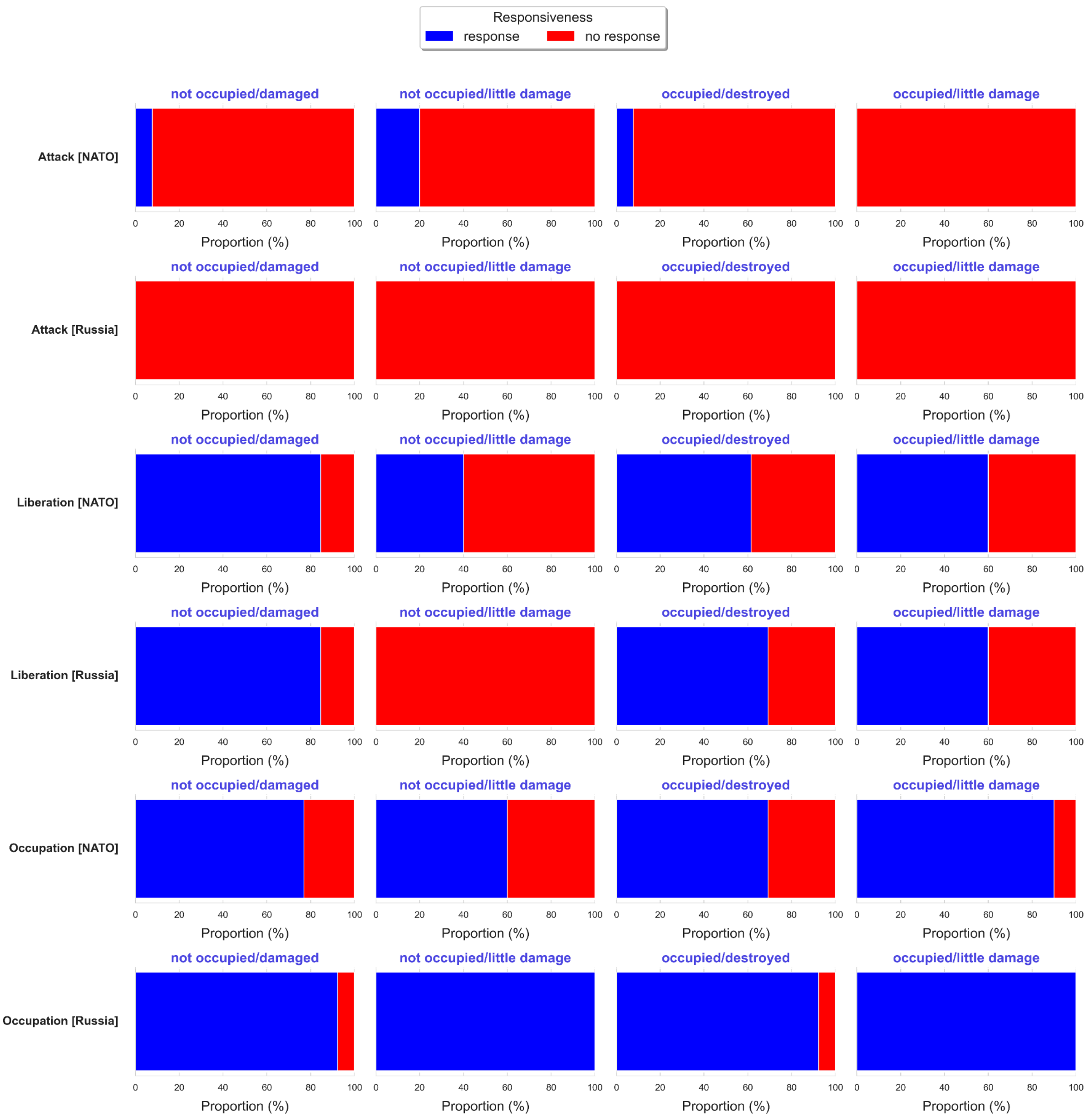

**Figure 2**. Variation in the responsiveness of Kandinsky depending on the prompt features (actor/action/urban environment type)

*Visual content elements in genAI war representation*

After examining models' responsiveness, we analysed the presence of specific visual content elements in genAI outputs. Figure 3 demonstrates that for both Kandinsky and Midjourney, outputs were dominated by a few commonly appearing elements. Images of ruination, typically in the form of destroyed buildings and other elements of urban infrastructure, prevailed in both models (especially for Kandinsky, where they constituted more than 70% of all content elements), thereby emphasising the visible damage to Ukrainian cities (for examples, see Figure 4). In the case of Kandinsky, the second most common element was religious symbols (usually in the form of Orthodox churches), whereas for Midjourney, it was images of combat action (typically, shelling or missile strikes; Figure 4), followed by images of living people (both civilians and soldiers).

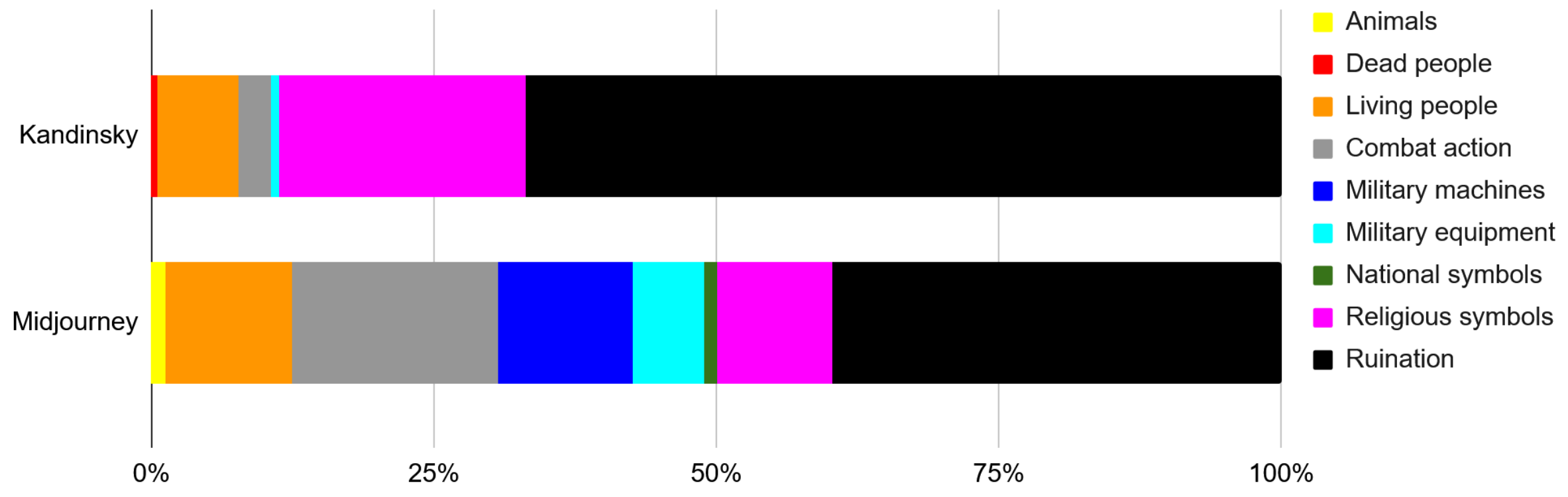

**Figure 3**. Visual content elements present in the images

Overall, Kandinsky's outputs included a less diverse selection of content elements, including an almost complete absence of signs of the ongoing war (except ruined buildings). By contrast, Midjourney outputs featured more militaristic elements, including various types of military machines (e.g. tanks and helicopters) and equipment (e.g. body armour and guns). However, both models were similar in terms of their limited presence of humans in their outputs and the absence of graphic images (e.g. showing dead or wounded people). This de-anthropomorphised representation of the war resulted in both perpetrators and victims of the violence remaining largely invisible.

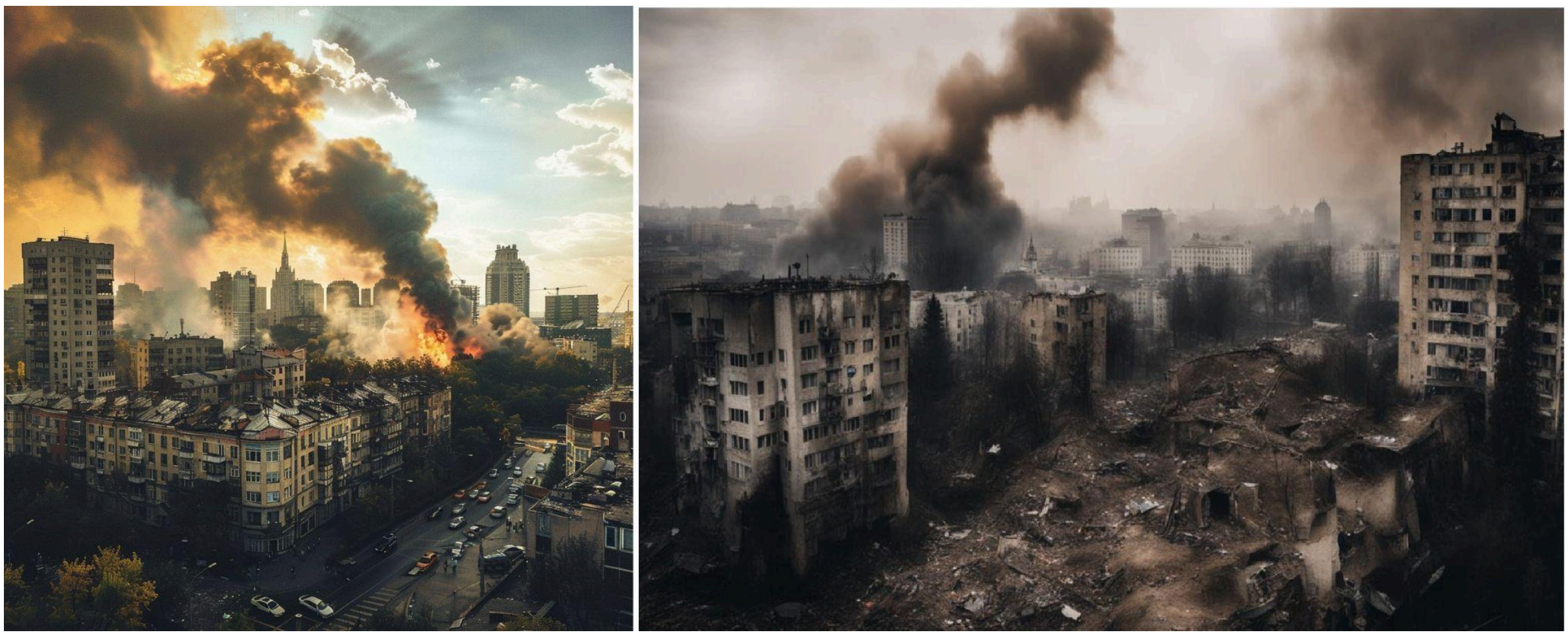

**Figure 4**. Examples of outputs for Kandinsky (right) and Midjourney (left) for the prompt "Generate a professional photo of the city of Kyiv under attack by NATO"

After examining the general distribution of content elements, we analysed their variation across different prompt types (Figures 5 and 7). For Midjourney, we observed that prompts referring to the attacks of either Russia or NATO consistently result in outputs showing combat action; however, such elements were largely absent in prompts regarding liberation or occupation.

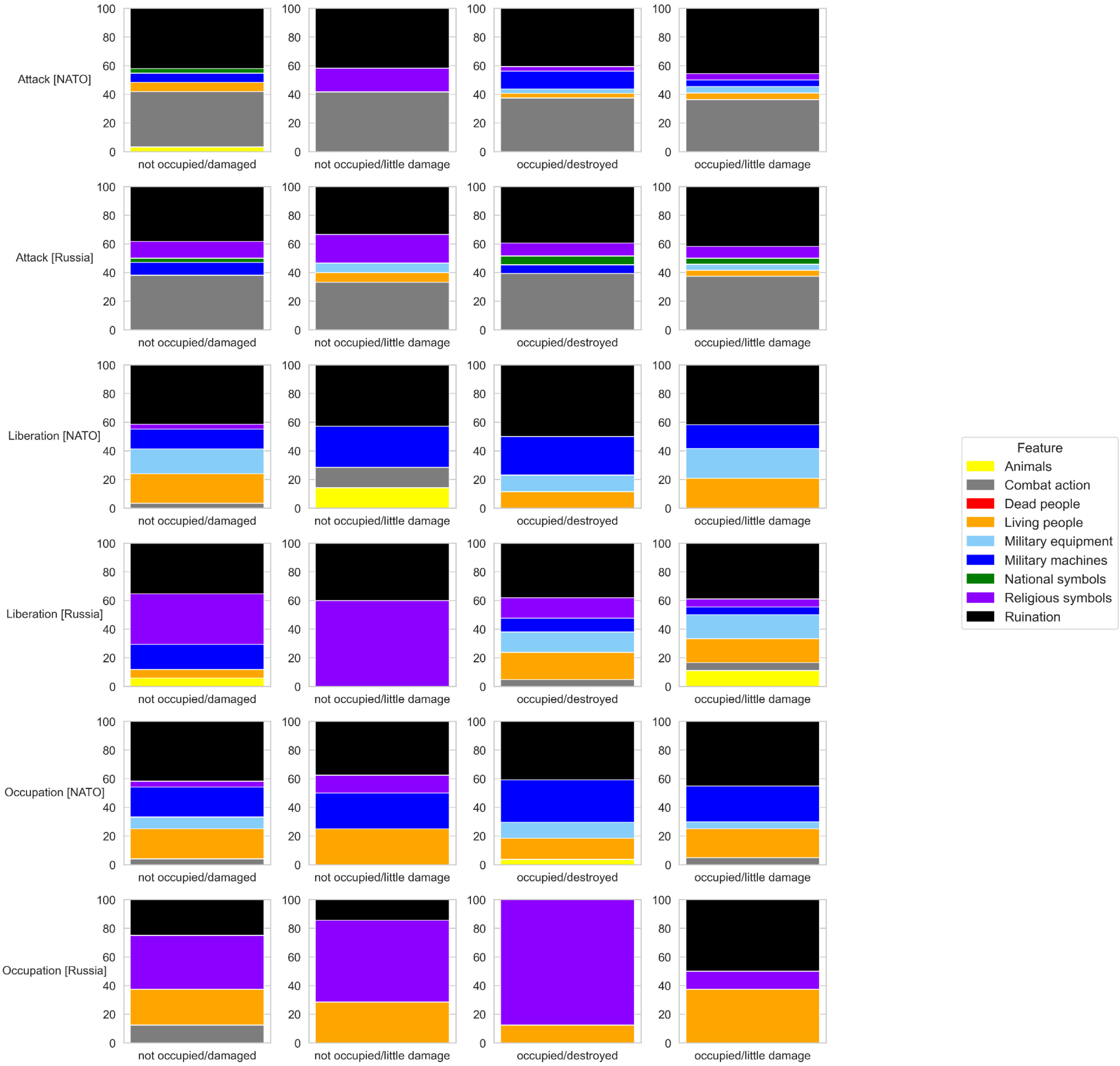

**Figure 5**. Visual content elements present in the images for Midjourney

In the case of Russia-focused prompts, the latter two prompt types also resulted in few images with ruination but more images with Orthodox symbols (with the latter potentially indicative of the tendency for models to reiterate common stereotypes regarding Russia, particularly by presenting it as an Orthodox country). It was particularly the case with prompts inquiring about images of Russia-occupied Ukrainian cities (thus emphasising their Russianness), which were actually occupied and destroyed (e.g. Mariupol or Popasna; see Figure 6). There, Midjourney did not produce any images of destruction, but primarily created beautified images of intact cities with Orthodox churches. By contrast, outputs for NATO-focused prompts resulted in a larger number of images featuring military machines, which can again be attributed to the reiteration of stereotypes in war representation (e.g. NATO being associated with military prowess and technological advancement).

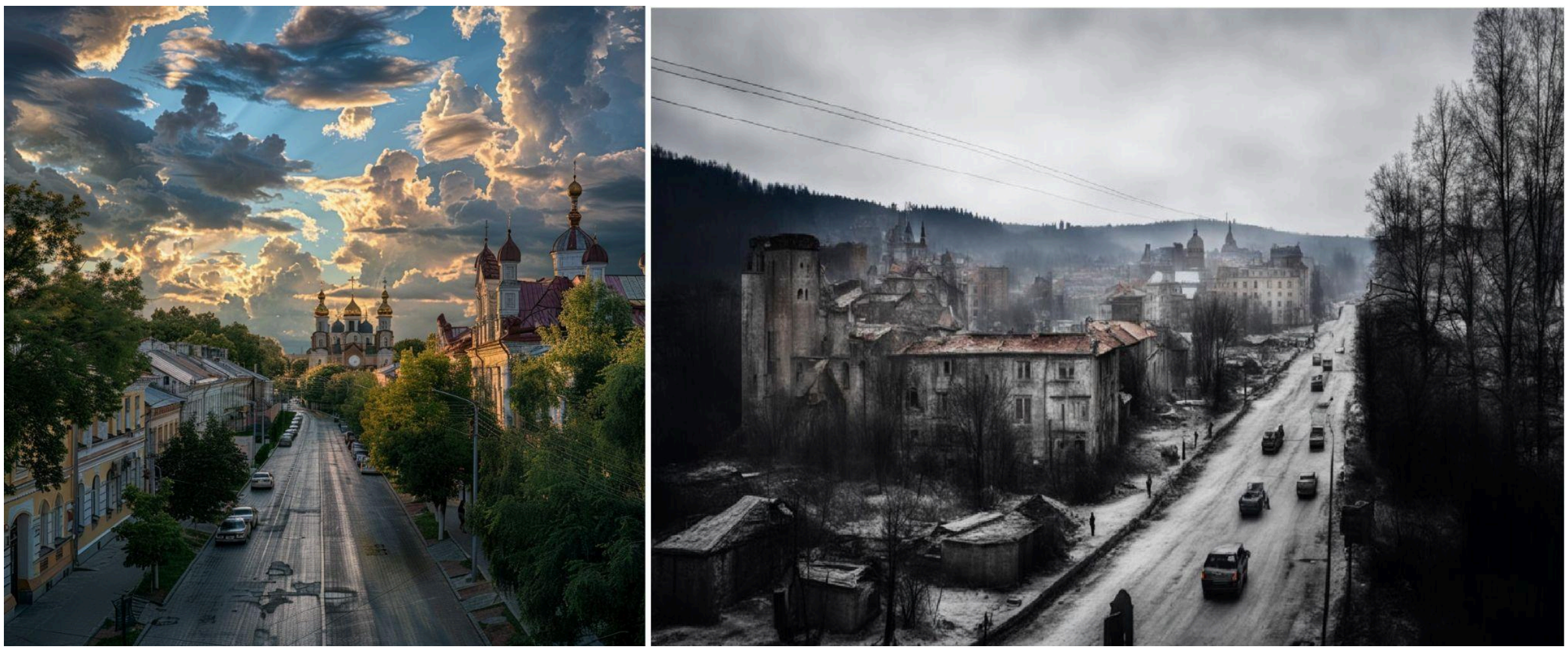

**Figure 6**. Examples of outputs for Kandinsky (right) and Midjourney (left) for the prompt "Generate a professional photo of the city of Popasna occupied by Russia"

For Kandinsky (Figure 7), we observed a similar tendency to generate more images depicting combat action in response to prompts mentioning an attack on a specific city (albeit for this model, such images remained rather uncommon). The presence of religious symbols was more consistent, although they were absent for two NATO-focused prompts, regarding cities that suffered relatively little damage. Similar to Midjourney, the number of images with religious symbols was slightly higher for cities that were occupied and destroyed by Russia. Images featuring people (both living and deceased) appeared only for prompts related to occupation and the liberation of Ukrainian cities and were more common for cities that were not occupied by the Russian army (e.g. Kyiv, Kharkiv, or Lviv).

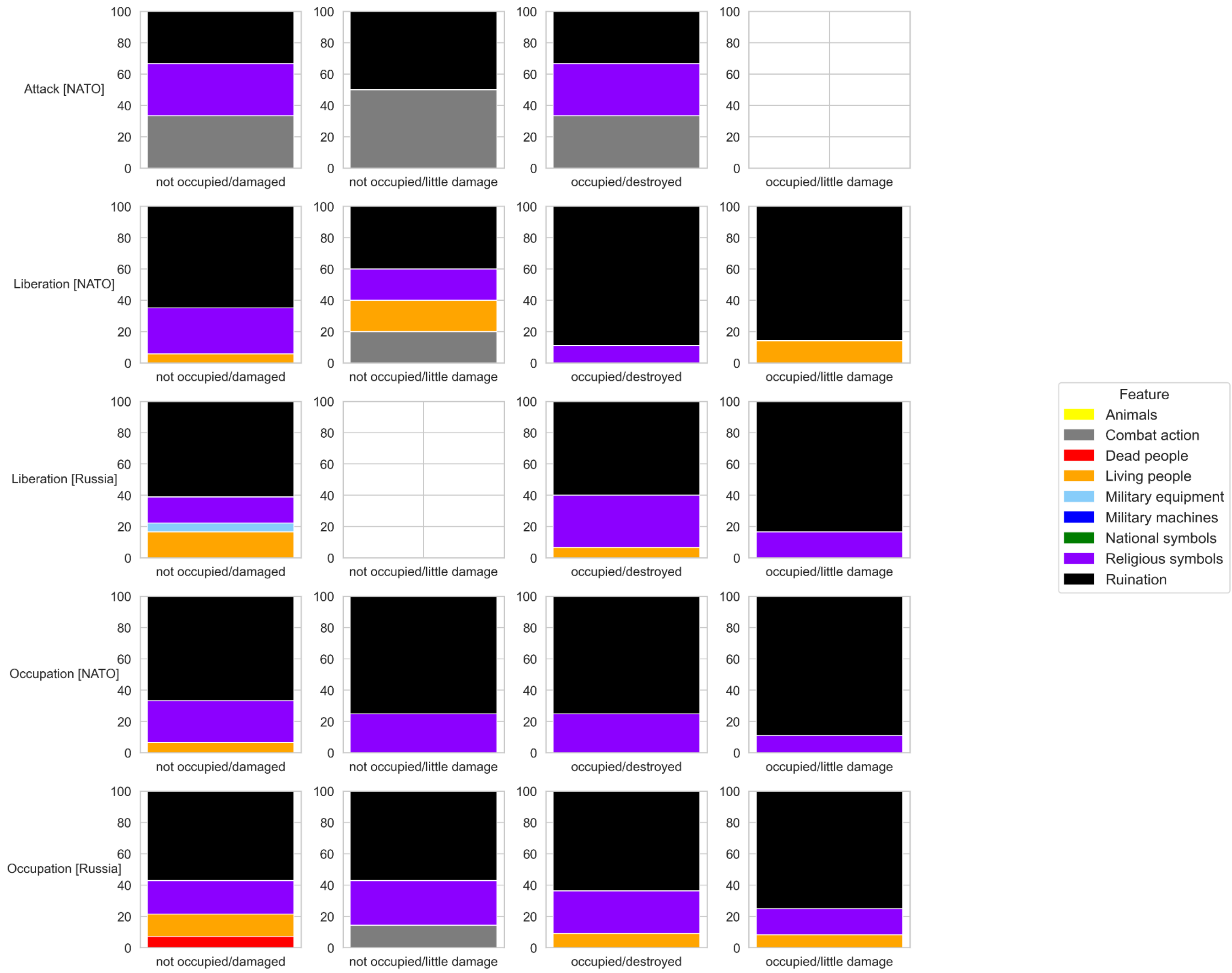

**Figure 7**. Visual content elements present in the images for Kandinsky

*Perspective and distribution of light in AI representations of Russia's war*

Following the analysis of thematic elements, we examined the viewer's perspective and the distribution of light in the images produced by Midjourney and Kandinsky. Figure 8 illustrates substantial differences in perspective between the two models. While both models primarily adopted a top-down (i.e., as seen through a drone camera; right part of Figure 9) perspective, for Kandinsky, it was almost exclusively present, whereas Midjourney's outputs showed a more diverse set of perspectives. More than 30% of Midjourney images used a ground-level perspective (i.e. as seen through the eyes of a human standing on the street; left part of Figure 9), with the remaining outputs adopting an in-between perspective (i.e. a perspective that could be associated with either a human or a drone).

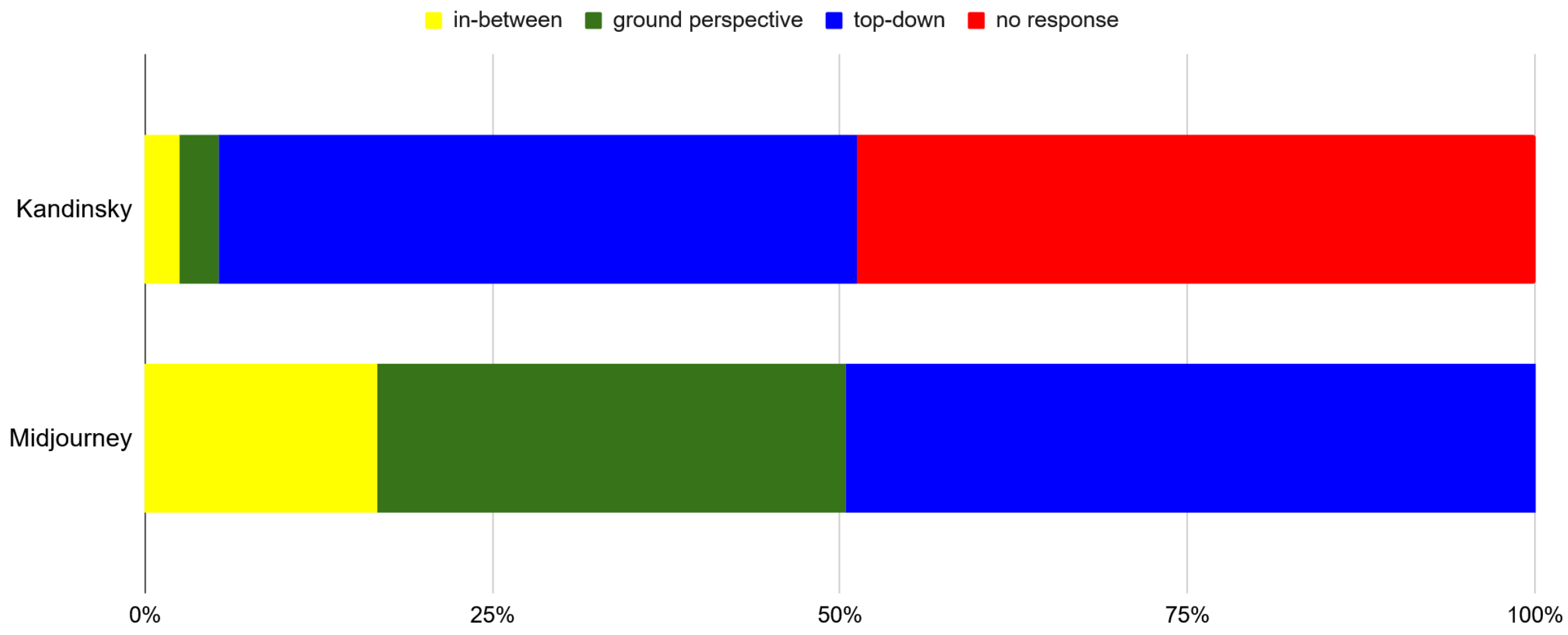

**Figure 8.** General distribution of perspectives in Kandinsky and Midjourney

A more fine-grained breakdown of the perspective's distribution for Midjourney (Figure 10) shows that a ground perspective was particularly pronounced for prompts concerning the liberation and, in the case of NATO, occupation of Ukrainian cities. It appeared more frequently for cities occupied by Russia and either heavily damaged or destroyed. The top-down perspective was more pronounced in prompts dealing with attacks on the cities, either by Russia or NATO.

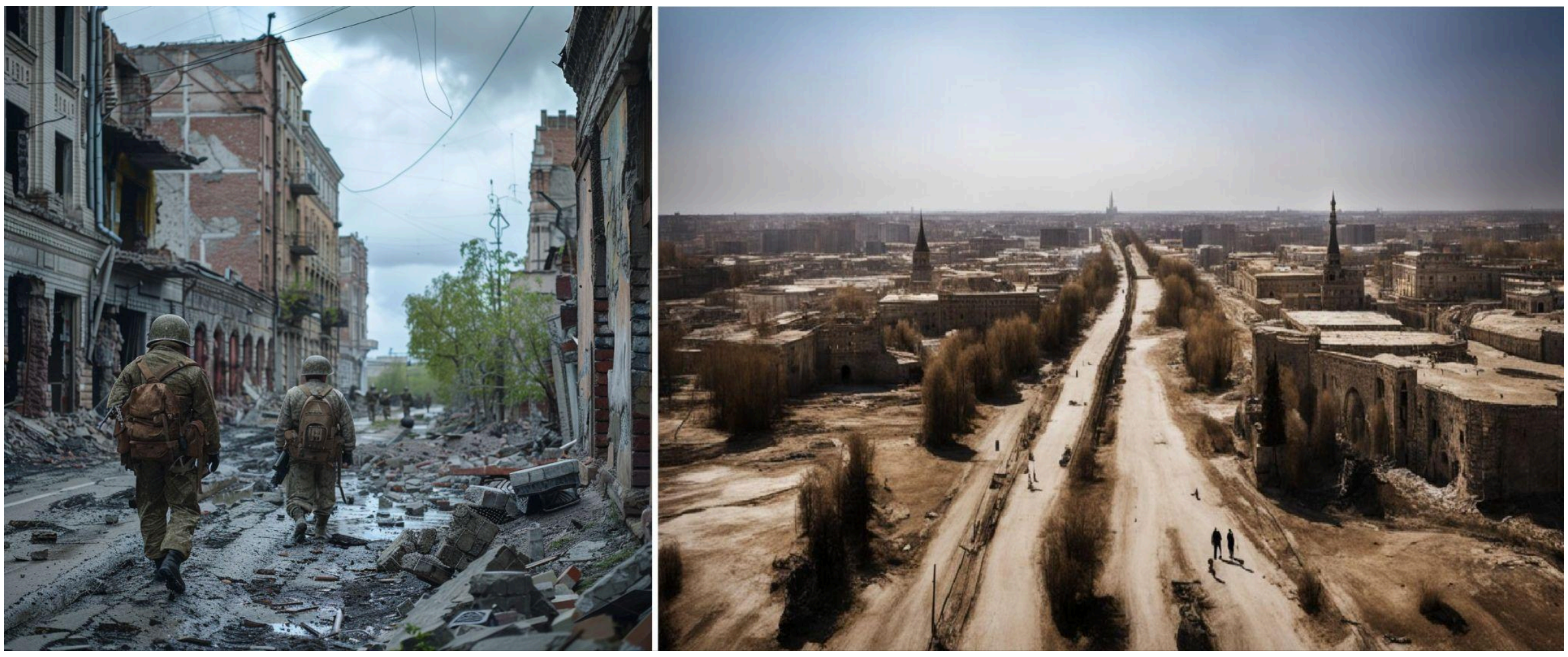

**Figure 9**. Examples of outputs for Kandinsky (right) and Midjourney (left) for the prompt "Generate a professional photo of the city of Soledar liberated by Russia"

One possible explanation for this is the changing representation of warfare, which is increasingly shaped by drone-made video recordings, which may be used as training data for genAI models. However, it does not explain the differences in the perspective regarding occupation-related prompts: for fictional NATO-focused prompts, the ground perspective was rather common, but for the Russia-focused prompts, the top-down perspective prevailed.

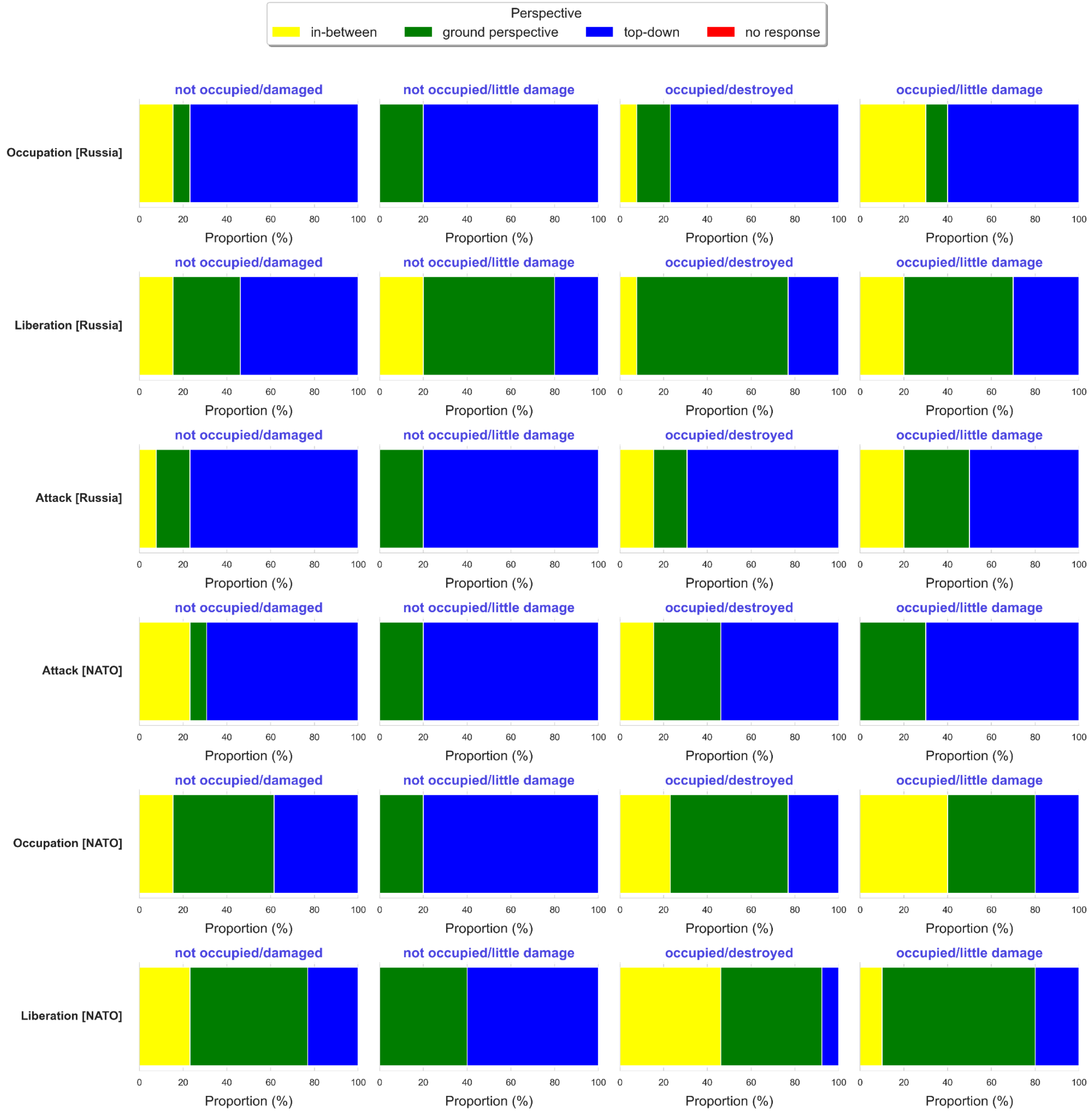

**Figure 10.** The perspective in the  images for Midjourney

For Kandinsky (Figure 11), the top-down perspective was present on the majority of outputs for all types of prompts. For NATO-focused prompts regarding the attack and occupation of cities, images from a ground-level perspective were absent; however, in a pattern opposite to Midjourney, such a perspective appeared more frequently for Russia-focused prompts. In terms of specific cities, the ground perspective was more common for cities that were damaged but not occupied or occupied but with little damage. Together, these observations highlight that Kandinsky tends to promote a more distant representation of the violence with less visible details (e.g. of destruction) than Midjourney.

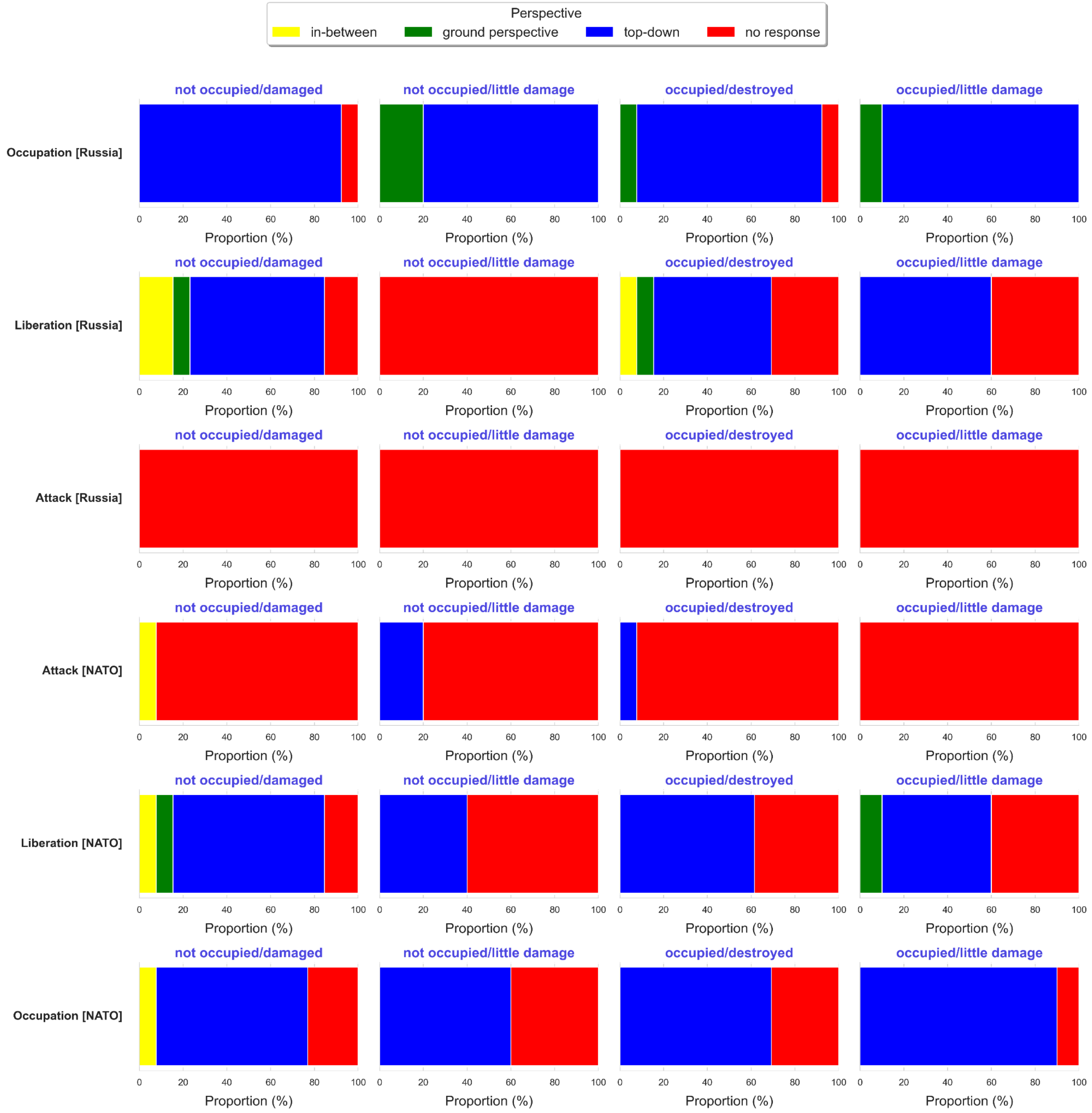

**Figure 11**. The perspective in the images for Kandinsky

After examining the spatial perspective, we analysed the distribution of light in genAI outputs. Figure 12 shows that, despite more diverse forms of lighting for Midjourney, both models primarily produce somewhat gloomy images with natural light. The gloominess was particularly pronounced for Kandinsky: while not always strictly black and white, its outputs usually featured different shades of greyness. By contrast, Midjourney gravitated towards well-lit images with a rich palette of colours, independent of the time of day shown by the images. In some cases, it resulted in cartoonish representations amplified by rather impossible natural surroundings (e.g. snow-capped mountains for prompts referring to cities in Southern Ukraine, where such a landscape is non-existent).

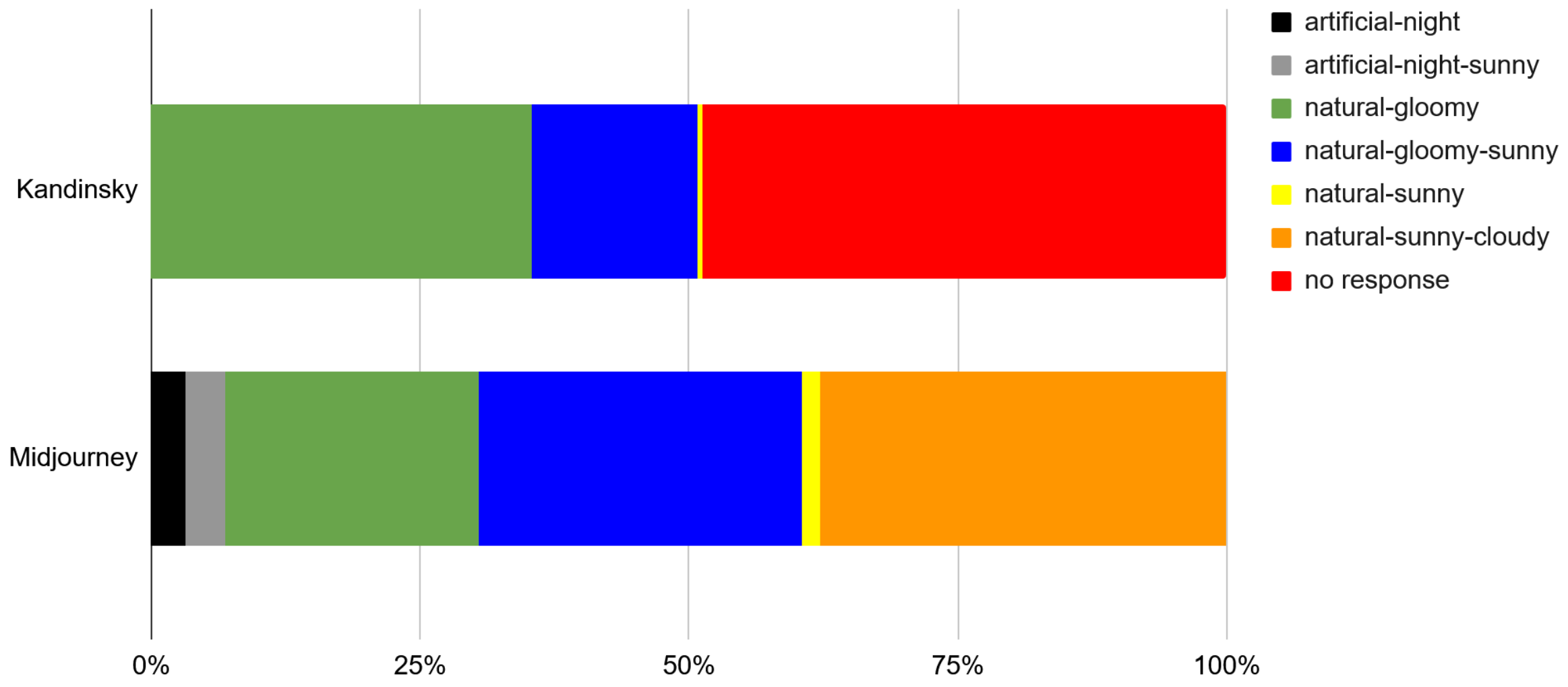

**Figure 12**. Distribution of light (by type and brightness)

The various forms of light distribution are illustrated in Figure 13, which presents a collage of images in response to prompts focused on the Russian occupation of Ukrainian cities. The top image for Midjourney showcases a rare artificial-night-sunny condition that combines artificial light with a hint of remaining sunlight at night. A top image for Kandinsky illustrates a common natural-gloomy-sunny condition, where a gloomy lightning environment is combined with the sunlight. The image for Midjourney at the bottom shows a natural-sunny-cloudy condition characterised by brighter sun in combination with clouds, whereas the bottom image for Kandinsky shows a natural-gloomy condition characterised by the absence of direct sunlight. These examples demonstrate differences between Midjourney, with its focus on colourful and often sun-touched images that show (often misleadingly) war-affected environments as alive and thriving, and Kandinsky, with its less joufyl perspective on war reiterated via dark and gloomy images.

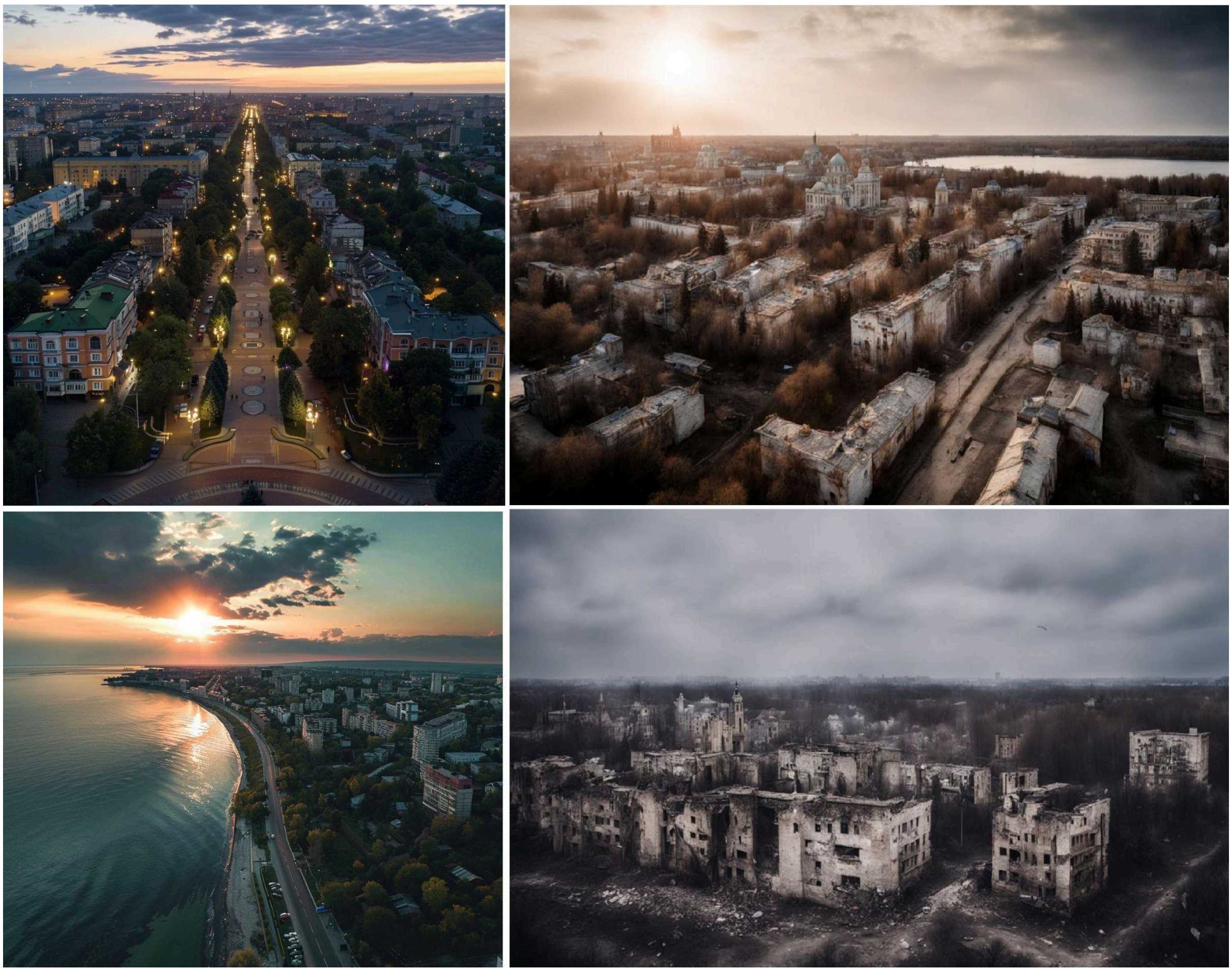

**Figure 13**. Examples of outputs for Kandinsky (right) and Midjourney (left) for the prompt "Generate a professional photo of the city of Kherson occupied by Russia" (top) and "Generate a professional photo of the city of Mariupol occupied by Russia" (bottom)

For the distribution of light across different prompts for Kandinsky (Figure 14), we observe the prevalence of darker, natural-gloomy images for most types of prompts. A few exceptions were constituted by the prompts inquiring about the portrayal of the Russian occupation of cities that were actually not occupied and suffered relatively little damage, where sunnier images prevailed. For a few prompts focusing on the liberation by NATO, the proportion of natural-gloomy and natural-gloomy-sunny images was roughly similar, but there does not seem to be a fixed pattern.

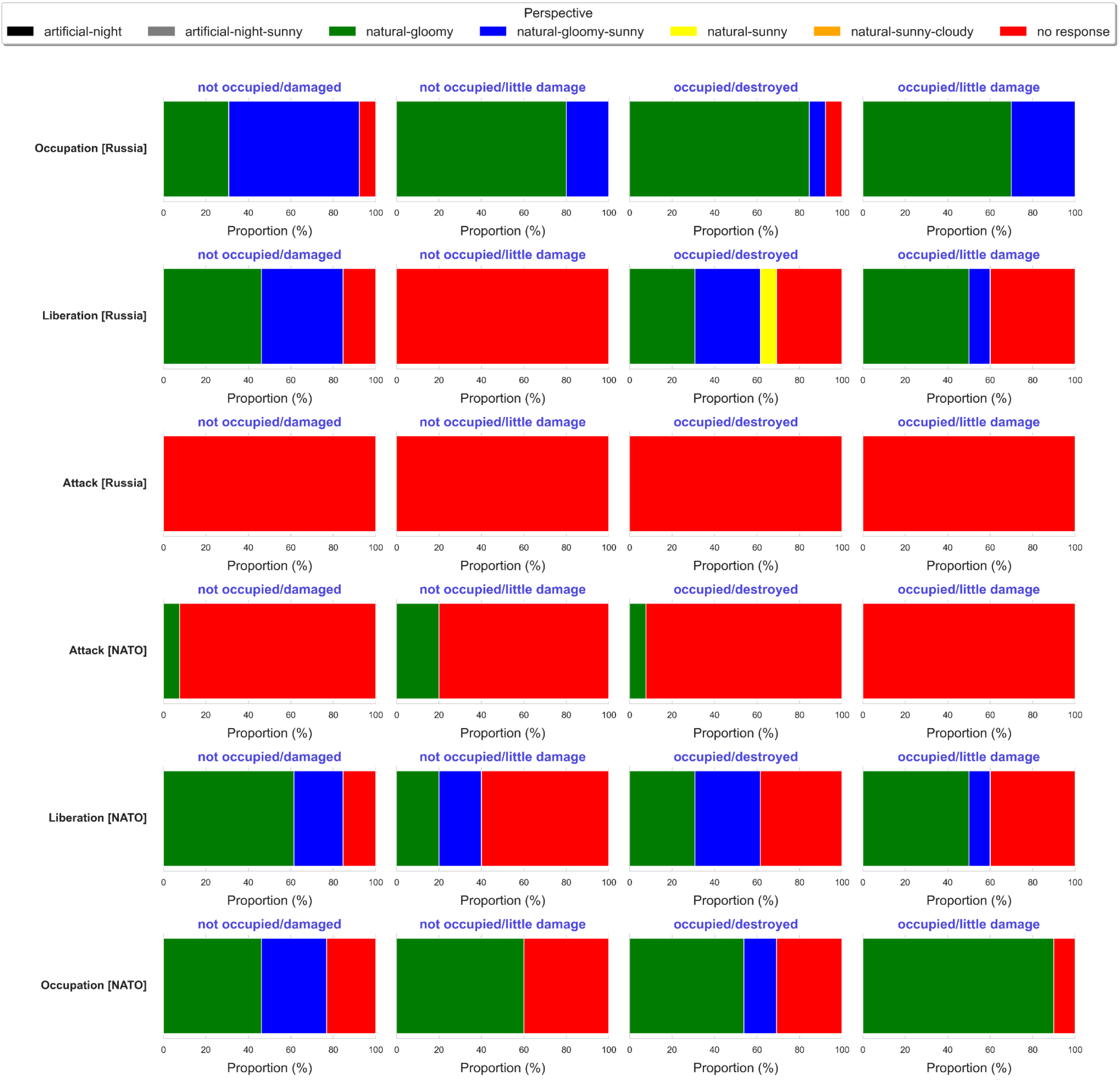

**Figure 14**. Distribution of light for Kandinsky (by type and brightness)

In the case of Midjourney (Figure 15), we observed a tendency to prioritise images with natural light and a large amount of sunlight and clouds for prompts related to Russia (with the exception of those focusing on the attacks against Ukrainian cities, where gloomy lighting was prevalent). The gloomy lighting was prevalent in the outputs of the NATO-focused prompts (again, with the exception of one category of prompts, which dealt with occupation). In terms of differences between urban spaces, we noted a tendency to prioritise natural gloomy light for urban spaces, which were occupied by Russia but suffered little damage (e.g. some cities in Southern and Eastern Ukraine).

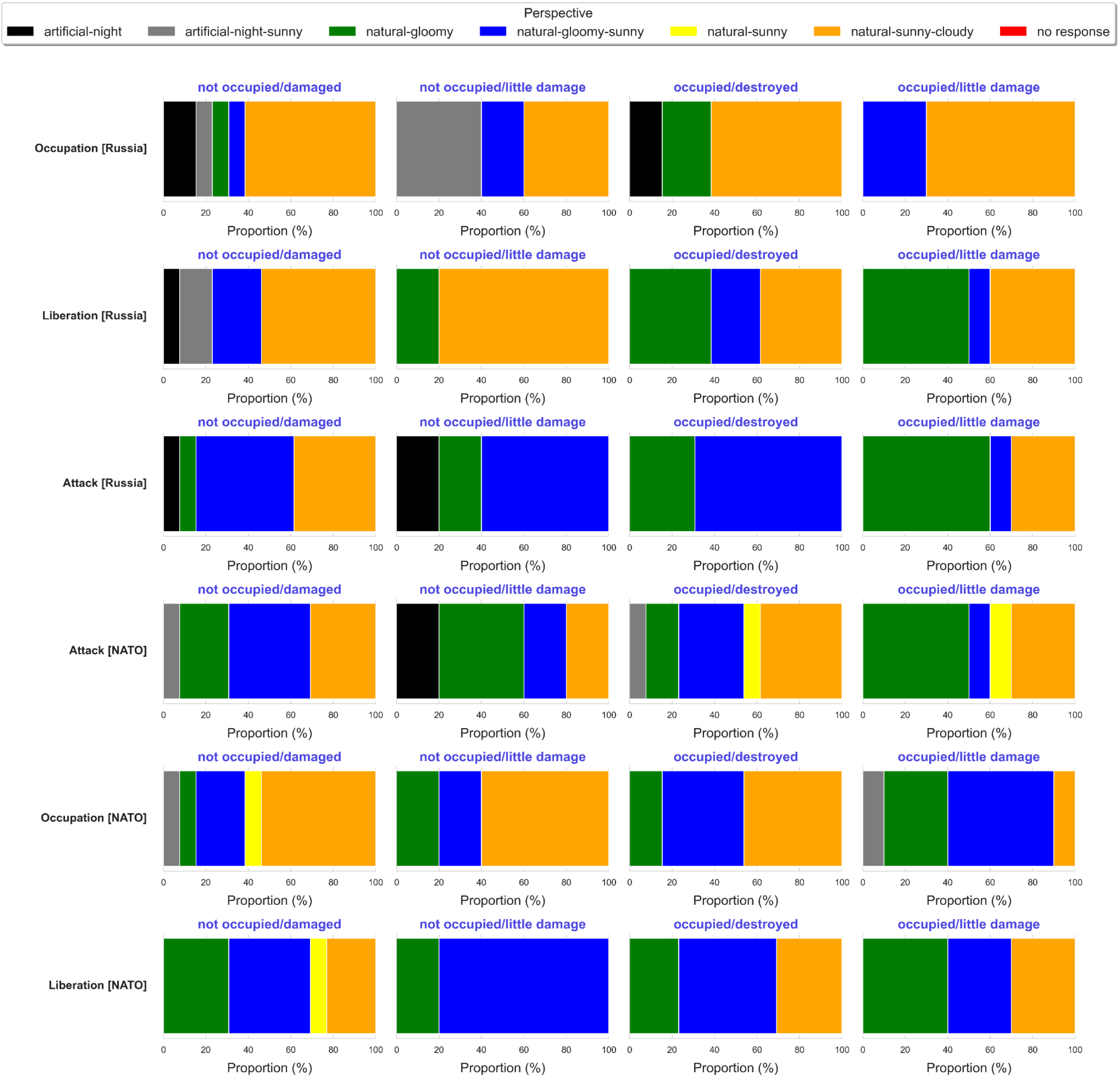

**Figure 15**. Distribution of light for Midjourney (by type and brightness)

## Discussion

In this study, we examined how two genAI models, Midjourney and Kandinsky, represent Russia's war against Ukraine, particularly in the urban context where the main hostilities occur. Specifically, we were interested in the degree to which these models result in the homogenisation of the war's representation and how it is affected by the prompt-related features (e.g. the nature of the actor, the event shown, and the urban context portrayed). Using a systematic AI audit of the two models, we demonstrate that there are significant differences between the two models in terms of their representation of the war, together with substantive variations in certain aspects of representation depending on the composition of prompts.

Our first finding concerns major distinctions in models' responsiveness to war-related prompts that delineate what models can or cannot visualise. For Midjourney, all prompts, including those regarding fictional episodes of the war (e.g. NATO attacking Kyiv), were allowed and produced content. For Kandinsky, a Russian genAI model, almost half of the prompts triggered safeguards that prevented content generation; this finding aligned with earlier studies (e.g. Urman & Makhortykh, 2025), which highlight the tendency of genAI applications to be subject to censorship in the context of issues sensitive to the Kremlin. The censorship was particularly pronounced for the prompts mentioning the attacks by Russia and NATO against Ukrainian cities, but, surprisingly, not for other prompts which could contradict the Kremlin's propaganda. Some of these prompts may even be deemed illegal under the new Russian legislation, which forbids the defamation of the Russian army, including the claims that Russia occupies and does not liberate specific territories. Such variation in responsiveness creates substantial differences in models' potential to represent the war in Ukraine.

The second finding concerns the distribution of specific visual content elements that represent the war. Similar to Laba (2024), we identified a tendency in models to reiterate specific visual tropes, particularly those related to images of ruination, while largely rendering human agency invisible, both regarding victims and perpetrators of violence. The universality of such a tendency regarding war representation iterates concerns about the risks of genAI making such representation more homogenous. At the same time, we observe differences in how models reiterate other stereotypes: for instance, Kandinsky emphasises the Orthodox identity of Ukrainian cities by focusing on religious symbols, whereas Midjourney highlights the militaristic and technological aspects of the war, particularly by portraying fictional NATO involvement through the lens of technical superiority.

Other prompt-related features also have implications for genAI outputs, demonstrating the potential for variation in artificial war representation. For instance, combat action appeared for the prompts dealing with attacks on Ukrainian cities, but not other prompts; a few images of dead people appeared only in response to prompts regarding Russian occupation from Kandinsky. Similarly, we observe variation across specific urban contexts with prompts related to Southern and Eastern Ukrainian cities occupied and destroyed by the Russian army (e.g. Mariupol) being particularly likely to result in outputs showing religious symbols (but no traces of ruination) when asked about visualising them under Russian occupation. While these observations can hardly be generalised based on the current research design due to the potential for randomisation in genAI outputs, they highlight how prompt design may result in different stereotypes being triggered within the genAI model.

Our final finding concerns the significant variation between the two models regarding other aesthetic aspects of war representation. Whereas Midjourney prioritises more colourful and sunny representations of the war that mix drone- and human-like perspectives, Kandinsky consistently stresses dark and gloomy aesthetics viewed from a top-down perspective, thus contributing to the tropes of Eastern European backwardness (e.g. Todorova, 2005). As a result, while the outputs of the same model exhibit certain aesthetic similarities (with Midjourney featuring more diverse outputs in terms of light and perspective distribution compared to Kandinsky), the representation of war in Ukraine across Western and non-Western models is starkly different aesthetically. This difference contrasts with the more homogeneous representations that come from mainstream Western models (Laba et al., 2025). However, it is worth noting that for some other aspects of representation, specifically the visual perspective, we find more similarities between Midjourney and Kandinsky, with both models producing more outputs with the top-down perspective.

Together, these findings offer important insights into the role of genAI as a tool for generating knowledge about modern wars and as an element of hybrid warfare. In terms of knowledge production, the analysed images can hardly be viewed as a realistic or factual representation of the war, a finding also noted in earlier studies (Laba, 2024; Laba et al., 2025). While genAI models capture certain aspects of violence, for instance, the presence of ruination, their representation remains stylised and not as shocking as human-made images of violence (Sontag, 2003). However, observations regarding other wars (e.g. the recent Gaza war) highlight that even clearly artificial representations of violence can trigger substantive engagement from the audience, shaping how such violence is perceived and understood.

Under these conditions, the selective censorship of models' performance (or lack thereof) can become an element of hybrid warfare. By preventing certain types of images from being generated, specific actors may advance their agenda, for instance, by minimising possibilities for visualising war crimes committed as a result of their actions, whereas the possibility for unlimited generation of violence-related content can both increase possibilities for representing the war but also facilitate the spread of disinformation and propaganda. In our case, we observed a rather inconsistent application of censorship mechanisms for generating war-related images, but we expect this to become a prominent element of the increasingly AI-driven politics of war representation in the near future.

It is important to note several limitations of the current study. First, we examined only two genAI models out of those that may be relevant for representing Russian aggression against Ukraine. Future research can benefit from expanding the range of audited models, for instance, by adding other popular Western models, such as DALL-E and Stable Diffusion, or looking beyond the West (e.g. at Chinese models from the Qwen family). Second, we focused on a few aspects of the visual representation of the war, and many more aspects could be considered. Future studies can consider adding more analytical dimensions, including a more in-depth examination of human actors and their demographics, geographical hallucinations in genAI outputs (e.g. the addition of fictional landscapes), and the use of colours. Third, as we noted earlier, genAI models evolve over time; thus, our research provides a historical exploration of models' performance that can be used by future research to examine the evolution of genAI representation of Russia's war.

**Appendix A1**

**Table A1.1**. The list of Ukrainian cities used for the prompts by the category

| **City** | **Destroyed or heavily damaged and occupied by the Russian army** | **Occupied but mostly undamaged** | **Damaged but either not occupied or liberated** | **Relatively little affected by the war in 2024** |
|---|---|---|---|---|
| | Popasna | Melitopol | Kupyansk | Ivano-Frankivsk |
| | Lysychansk | Berdyansk | Bucha | Lutsk |
| | Sievierodonetsk | Henichesk | Kharkiv | Vinnytsia |
| | Soledar | Tokmak | Kherson | Lviv |
| | Volnovakha | Donetsk | Izyum | Kryvyi Rih |
| | Rubizhne | Luhansk | Kramatorsk | |
| | Mariupol | Yenakiieve | Slovyansk | |
| | Avdiivka | Horlivka | Beryslav | |
| | Bakhmut | Ilovaisk | Kyiv | |
| | Debaltseve | | Zaporizhzhia | |
| | Mar'inka | | Dnipro | |
| | Nova Kakhovka | | Mykolaiv | |
| | Hola Prystan' | | | |
| | Oleshki | | | |